\documentclass[aps]{revtex4}
\input{epsf}
\def\beq{\begin{eqnarray}}
\def\eeq{\end{eqnarray}}
\def\nn{\nonumber\\}
\def\half{{\textstyle{1\over2}}}
\begin{document}
\title{Counting of Black Hole Microstates}
\author{A. Ghosh}\thanks{amit.ghosh@saha.ac.in} 
\author{P. Mitra}\thanks{parthasarathi.mitra@saha.ac.in}
\affiliation{Theory Division, Saha Institute of Nuclear Physics\\
1/AF Bidhannagar, Calcutta 700064}
\begin{abstract}
The entropy of a black hole can be obtained by counting states
in loop quantum gravity. The dominant term depends on the Immirzi
parameter involved in the quantization and 
is proportional to the area of the horizon, while there
is a logarithmic correction with coefficient -1/2.
\end{abstract}
\maketitle
\vskip .5in
%\rightline{gr-qc/0603029}
%%%%%%%%%%%%%%%%%%%%%%
%%%%%%%%%%%%%%%%%%%%%%

It is an honour and a pleasure to write in the volume dedicated to Professor Amal Kumar
Raychaudhuri, eminent theoretical physicist and revered teacher of generations of Physics
students. The theory of gravitation, with which he preoccupied himself, is progressing steadily,
and although a full quantum theory is not yet at hand, a lot of interesting results are
available.

A framework for the description of quantum gravity using holonomy variables has become
popular as loop quantum gravity \cite{ash}. A start was made in this work in the direction
of counting of black hole microstates. Further progress was made in \cite{gm0}, \cite{meissner}
and in \cite{gm}. In the present article we shall try to tie up some loose ends left
there. Other discussions of the subject can be found in \cite{tamaki,0511084}. 

In this approach, there is a classical isolated horizon and quantum states are
sought to be built up by associating spin variables with punctures on the
horizon. The entropy is obtained by counting the possible states that are
consistent with a particular area, or more precisely with a particular
eigenvalue of the area operator \cite{ash}.

%\section{Counting of states}

We set units such that $4\pi\gamma\ell_P^2=1$, where $\gamma$ is the Immirzi
parameter and $\ell_P$ the Planck length. Equating the classical area $A$ of the horizon
to the eigenvalue of the area operator we find
\beq A=2\sum_{p=1}^N\sqrt{j_p(j_p+1)}\;,\label{areacons}\eeq
where the $p$-th puncture carries a spin $j_p$, more accurately an irreducible representation
labelled by $j_p$, and contributes a {\em quantum of area} $2\sqrt{j_p(j_p+1)}$ to the total
area spectrum. For mathematical convenience let us replace the half-odd integer spins
by integers $n_p=2j_p$, which makes the area equation $A=\sum_p\sqrt{n_p(n_p+2)}$. Henceforth,
$n_p$ will be referred to as the {\em `spin'} carried by the $p$-th puncture. A puncture carrying
zero spin contributes nothing to the spectrum, hence such punctures are irrelevant. Since the
minimum `spin' each puncture should carry is unity the total number of punctures cannot exceed
$A/\sqrt 3$. At the same time, the largest `spin' a puncture can carry is also bounded,
$n\leq N$, where $\sqrt{N(N+2)}=A$.

A sequence of `spins' $n_p$, each $1\leq n_p\leq N$, will be called {\em permissible}
if it obeys (\ref{areacons}). The $p$-th puncture gives $(n_p+1)$ number of quantum states. In
this way each permissible sequence gives rise to a certain number of quantum states. The task is
to find the total number of states for all permissible sequences. Let it be $d(A)$. One can
subdivide the problem as follows: Fix {\em any} puncture, say $p=1$. Consider the subset of
all permissible sequences such that puncture 1 carries `spin' 1. For such sequences the area
equation (\ref{areacons}) reads
\beq \sum_{p\neq 1}\sqrt{n_p(n_p+2)}=A-\sqrt 3\;.\label{areacons1}\eeq
So the total number of quantum states given by all sequences obeying (\ref{areacons1}) is
$d(A-\sqrt 3)$. But the puncture 1 itself gives two states. Therefore, the total number
quantum states given by the subset of permissible sequences in which puncture 1 carries `spin' 1
is $2d(A-\sqrt 3)$. In the next step consider the subset of all permissible sequences such
that puncture 1 carries `spin' 2. Arguments similar to the above leads to the total number of states
for such subset of sequences as $3d(A-2\sqrt 2)$. Continuing this process we end up with a
recurrence relation
\beq d(A)=\sum_{n=1}^{N-1}(n+1)d(A-\sqrt{n(n+2)})+N+1.\label{recr}\eeq
This is similar to the relation in \cite{meissner}, but differs from it in
having all values of $m=-j,-j+1,...+j$ allowed.

In solving (\ref{recr}) we employ a trial solution $d(A)=\exp(\lambda A)$. Then (\ref{recr})
puts a condition on $\lambda$:
\beq \sum_{n=1}^N(n+1)\,e^{-\lambda\sqrt{n(n+2)}}=1\;. \label{lconst1}\eeq
Therefore, a solution for $\lambda$ obeying the above equation implies a solution of
the recurrence relation (\ref{recr}). For large area $A\gg 1$, we have $N\gg 1$. Moreover, for
$\lambda=o(1)$ the summand falls off exponentially for large $n$. So formally we can extend
the sum up to infinity. This numerically yields $\lambda\simeq 0.861$. The error we make in
estimating $\lambda$ by extending the sum all the way to infinity is $o(e^{-A})$. The total
degeneracy $d(A)$ then gives rise to a Boltzmann entropy $S(A)=\ln d(A)=\lambda A$. In
physical units
\beq S(A)={\lambda A\over 4\pi\gamma\ell_P^2}\;,\eeq
which yields $A/4\ell_P^2$ if we choose the parameter $\gamma=\lambda/\pi$. This is the basic
idea behind the counting and thereby, making a prediction for the $\gamma$-parameter in order
that an entirely quantum geometric calculation matches a semi-classical formula. Thus we
cannot derive the semiclassical world but can adjust our parameters in the theory such that
the semiclassical world emerges.

In the above counting process we completely miss which configuration of spins dominates the
counting, in other words contributes the largest number of quantum states. A common
misconception is that the smallest `spin' $n=1$ at every puncture gives rise to the largest
number of quantum states. It arises from the intuition that such configuration maximizes the
number of punctures and is therefore semiclassically favoured. The following analysis will show
that such an intuition is incorrect. We focus on punctures carrying identical spins. This is
somewhat in analogy with statistical mechanics where we look for particles carrying the same
energy. Let the number of punctures carrying `spin' $n$ be $s_n$. So in the area equation
(\ref{areacons}) the sum over punctures can be replaced by the sum over spins
\beq A=\sum_ns_n\sqrt{n(n+2)}\;.\label{areacon2}\eeq
Equation (\ref{areacon2}) further symbolizes the fact that spins are more fundamental in this
problem than punctures. A configuration of `spins' $s_n$ will be called permissible if it obeys
(\ref{areacon2}). Each configuration yields $\prod_n(n+1)^{s_n}$ quantum states but
each of the configurations can be chosen in $(\sum s_n)!/\prod s_n!$ ways (punctures
are considered distinguishable). Therefore, the total number of quantum states given by such a
configuration is
\beq d_{s_n}={(\sum_ns_n)!\over\prod_ns_n!}\prod_n(n+1)^{s_n}\;.\label{sconf}\eeq
However, the configuration in (\ref{sconf}) may not be permissible. To obtain a permissible
configuration we maximize $\ln d_{s_n}$ by varying $s_n$ subject to the constraint
(\ref{areacon2}). In the variation we assume that $s_n\gg 1$ for each $n$ (or only such
configurations dominate the counting). Such an assumption clearly breaks down if $A\sim o(1)$.
The variational equation $\delta\ln d_{s_n}=\lambda\delta A$, where $\lambda$ is a Lagrange
multiplier, gives
\beq {s_n\over\sum s_n}=(n+1)\,e^{-\lambda\sqrt{n(n+2)}}\;.\label{snsol1}\eeq
Clearly, for consistency, $\lambda$ obeys (\ref{lconst1}) with $N=\infty$ 
(cf. \cite{krasnov}). As already
observed this hardly makes a difference, more precisely the differences are exponentially
suppressed $o(e^{-A})$ for large areas. Moreover, although each $s_n\gg 1$, the sum $\sum s_n$
is convergent, since large $n$ terms are exponentially suppressed. This can be explicitly seen
by plugging in (\ref{snsol1}) into (\ref{areacon2}), which yields
\beq \sum s_n=A\left[\sum(n+1)\sqrt{n(n+2)}\,e^{-\lambda\sqrt{n(n+2)}}\right]^{-1}=
0.342A\;.\eeq

Plotting (\ref{snsol1}) we find how the configuration contributing the largest number of
quantum states is distributed over all spins. The maximum number of punctures carry integer 1
(which are truly spin 1/2), as the intuition suggests, but surprisingly all other spins 
also contribute.

\begin{figure}[h]\label{plot1}
\epsfxsize=2in \centerline{\epsffile{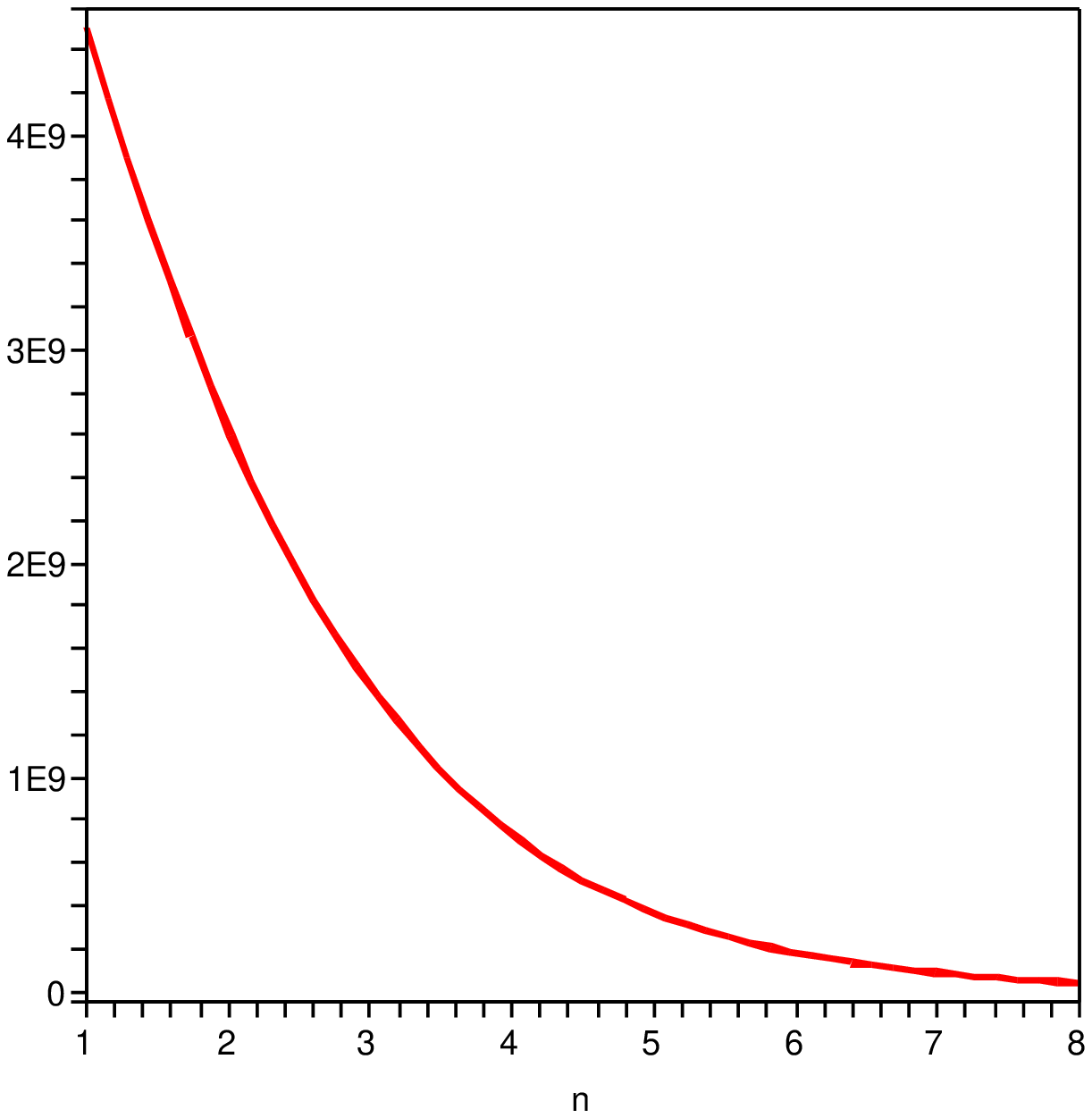}}
\caption{\em\baselineskip=11pt Dominant configuration.}
\end{figure}

Let us denote the configuration (\ref{snsol1}) dominating the counting by $\bar s_n$. The
total number of quantum states is obviously $d=\sum_{s_n}d_{s_n}$ where the sum extends over
all permissible configurations. However, the largest number of states come from some dominant
configuration $\bar s_n$. So we can expand $d$, more accurately the entropy $\ln d$, around
this dominant configuration and the result should be expressible in the form $\ln d=\ln
d_{\bar s_n}-\half\sum_{n,n'}\delta s_nK_{nn'}\delta s_{n'}+o(\delta s_n^2)$, where $\delta
s_n$ satisfies the area equation $\sum\delta s_n\sqrt{n(n+2)}=0$, which follows by requiring that
the displaced configuration $\bar s_n+\delta s_n$ also obeys the area equation
(\ref{areacon2}). One may wonder at this point whether such a condition can ever be met since
$\delta s_n$ are integers whereas $\sqrt{n(n+2)}$ are irrational. Strictly speaking in the
area equation (\ref{areacon2}) we require that the sum $\sum s_n\sqrt{n(n+2)}$ should be close
to $A$. In other words a range $-\Delta\leq A\leq\Delta$, where $\Delta\ll A$, must
exist such that the sum lies in the range. This amounts to saying that $\sum\delta
s_n\sqrt{n(n+2)}$ be a number $\epsilon\sim o(1)$, where $\epsilon$ may vary with
configurations but the variation is slow. The matrix $K_{nn'}$ which depends on $\bar s_n$ is
symmetric. A simple calculation gives $K_{nn'}=\delta_{nn'}/\bar s_n-1/(\sum \bar s_m)$.
The total number of states can be expressed as
\beq d=d_{\bar s_n}\sum_{-\infty}^\infty e^{-\half\sum_{n,n'}\delta s_nK_{nn'}\delta
s_{n'}}\delta(\sum\delta s_n\sqrt{n(n+2)})\label{sum1}\eeq
where the sum extends over all fluctuations. The large fluctuations die out exponentially. 
The Gaussian sum over fluctuations would have produced a factor $1/\sqrt{{\rm det}(K)}$ if the
delta function were not there. It is easy to see that $K$ has a zero eigenvalue
($\sum K_{nn'}\bar s_{n'}=0$), so this hypothetical factor would be divergent.
But the delta function makes the sum over the zero mode of $K$ finite. 
Note that each 
nonzero eigenvalue of $K$ scales like $1/A$, so the fluctuations $\delta s_n$,
which have to be rewritten in terms of normal modes $\delta s_n'$ of $K$,
have to be converted to $\delta s_n'/\sqrt A$, producing extra factors of $\sqrt A$
for each summation. As one summation is removed by the delta function,
\beq d=Cd_{\bar s_n}\Big[\prod_n\sqrt A\,\Big]/\sqrt A,\eeq
where $C$ does not involve $A$.
Plugging (\ref{snsol1}) into (\ref{sconf}) and neglecting $o(1)$ factors we find 
\beq d_{\bar
s_n}=\exp(\lambda A)[\sum\bar s_n]^{1/2}/\prod_n(2\pi\,\bar s_n)^{1/2}.\eeq 
Noting that the factors of $\sqrt A$ cancel,  
we get $d=\exp(\lambda A)$ up to factors of $o(1)$ which will
anyway be of $o(1)$ in the entropy and therefore have been neglected throughout in the
calculation.

The above steps illustrate the basic points of the calculation which now can be adapted
to the actual counting. The actual counting problem involves another crucial condition: Each
puncture carrying a representation labelled by `spin' $n$ must be associated with a state $|n,m
\rangle$ where $m$ is half-odd integer valued spin projections, $-n/2\leq m\leq
n/2$. The condition is that $\sum_pm_p=0$ where the sum extends over all punctures. Therefore,
a sequence of `spins' $n_p$ is permissible if it obeys the area equation and the spin projection
equations simultaneously. The task is to count the number of states for all such permissible
sequences. A recurrence relation, similar to (\ref{recr}), can be found also in this case.
Following \cite{meissner} we relax the spin-projection equation to $\sum_pm_p=\nu$ where $\nu$
is a half-odd integer that can take any sign. Let the total number of states be $d_\nu(A)$. As
before fix a puncture, say 1, and let it carry `spin' 1. For such sequences the area and the
spin projection equations become $\sum_{p\neq 1}\sqrt{n_p(n_p+2)}=A-\sqrt 3$ and $\sum_pm_p=
\nu\pm 1/2$ respectively. Therefore, the number of quantum states for all permissible configurations such
that the puncture 1 carries `spin' 1 is $d_{\nu+1/2}(A-\sqrt 3)+d_{\nu-1/2}(A-\sqrt 3)$.
Continuing this process as before we end up with the recurrence relation
\beq d_\nu(A)=\sum_{n=1}^{N-1}\sum_{m=-n/2}^{n/2}d_{\nu-m}(A-\sqrt{n(n+2)})+1,\label{recr2}\eeq
where the largest `spin' $N$ contributes only one state to the above sum,
provided $\nu$ belongs to the set of allowed values of $m=\{-N/2,...N/2\}$. In order to solve
(\ref{recr2}) we consider the Fourier transform of $d_\nu(A)$
\beq d_\nu(A)=\int_{-2\pi}^{2\pi}{d\omega\over 4\pi}\;d_\omega(A)\,e^{i\omega\nu}\eeq
and re-express the recurrence relation in terms of $d_\omega(A)$:
\beq d_\omega(A)=\sum_n d_\omega(A-\sqrt{n(n+2)})\sum_m\cos(m\omega)\;.\label{recr1}\eeq
In an attempt to solve (\ref{recr1}) we again employ a trial solution $d_\omega(A)=\exp(\lambda
_\omega A)$, which on being plugged into the recurrence relation yields a condition on $\lambda_\omega$:
\beq 1=\sum_ne^{-\lambda_\omega\sqrt{n(n+2)}}\sum_m\cos(m\omega)\;.\label{newr}\eeq
The above equation (\ref{newr}) clearly shows that $\lambda_\omega$ is a periodic function of
$\omega$. It is also multi-valued. However, it has a local maximum at $\omega=0$ and in a small
neighbourhood of this maximum it can be approximated by a power series $\lambda_\omega=\lambda
+a_2\omega^2+a_4\omega^4+\cdots$. Values of $\lambda, a_2, a_4,...$ etc can then be obtained
from (\ref{newr}) by comparing various powers of $\omega$. It can be easily shown that $\lambda$
obeys the same recurrence relation as (\ref{recr}), therefore the same as before,
\beq a_2=-{\sum e^{-\lambda\sqrt{n(n+2)}}\sum{1\over 2}m^2\over\sum
(n+1)\sqrt{n(n+2)}e^{-\lambda\sqrt{n(n+2)}}}\simeq-0.151\;.\eeq
Finally, we are interested in
\beq d_0(A)=\int_{-2\pi}^{2\pi}{d\omega\over 4\pi}\;e^{\lambda_\omega A}={\alpha\over\sqrt
A}\;e^{\lambda A}\;,\quad{\rm where}\;\alpha\sim o(1)\;,\eeq
which yields an entropy $S(A)=\lambda A-{1\over 2}\ln A$, or 
\beq S(A)={\lambda A\over 4\pi\gamma\ell_P^2}
-{1\over 2}\ln A\label{log}\eeq
in physical units. Thus, incorporation of the projection
equation $\sum m=0$ does not alter the leading expression of entropy, hence does not give a
different requirement on the $\gamma$-parameter  to make the leading entropy agree with the semiclassical
formula, but gives a universal log-correction to the semiclassical formula with a factor of
$1/2$.

The counting using the dominant configuration is cleaner when the projection equation is
incorporated. Here we present the detailed calculation. As before, let $s_{n,m}$ denotes the
number of punctures carrying `spin' $n$ and projection $m$. The area and spin projection
equations take the form
\beq A=\sum_{n,m}s_{n,m}\sqrt{n(n+2)}\;,\quad 0=\sum_{n,m}ms_{n,m}\;.\label{newc}\eeq
A configuration $s_{n,m}$ will be called permissible if it satisfies both of these equations
(\ref{newc}). Since now the $m$-quantum numbers are also specified each puncture is in a
definite quantum state specified by two quantum numbers $n,m$. 
The total number of quantum states for all configurations is
the number of ways a configuration can be chosen. This can be done in two steps. Note that
$\sum_ms_{n,m}=s_n$. So first, the configuration $s_n$ can be chosen in $(\sum s_n)!/\prod
s_n!$ ways. Then out of $s_n$ the configuration $s_{n,m}$ can be chosen in $s_n!/\prod_m
s_{n,m}!$ ways and finally a $\prod_n$ has to be taken. Thus
we get
\beq d_{s_{n,m}}={(\sum_ns_n)!\over\prod_ns_n!}\prod_n{s_n!\over\prod_ms_{n,m}!}={(\sum_{n,m}
s_{n,m})!\over\prod_{n,m}s_{n,m}!}\;.\label{newco}\eeq
To obtain permissible configurations which contribute the largest number of quantum states
we maximize $\ln d_{s_{n,m}}$ by varying $s_{n,m}$ subject to the two conditions (\ref{newc}). The
calculation is identical as before and the result can be expressed in terms of two Lagrange
multipliers $\lambda,\alpha$
\beq {s_{n,m}\over\sum s_{n,m}}=e^{-\lambda\sqrt{n(n+2)}-\alpha m}\;.\label{news}\eeq
Consistency requires that $\lambda$ and $\alpha$ be related to each other as $\sum_n
e^{-\lambda\sqrt{n(n+2)}}\sum_me^{-\alpha m}=1$. In order that (\ref{news}) satisfy the spin
projection equation we must require the sum $\sum_ne^{-\lambda\sqrt{n(n+2)}}\sum_mme^{-\alpha
m}=0$. This is possible if and only if $\sum_mme^{-\alpha m}=0$ for all $n$, which essentially implies
$\alpha=0$. (The value $2i\pi$ is excluded by positivity requirements.) 
Therefore, the condition on $\lambda$ becomes the same as before. The sum $\sum
s_{n,m}=\sum s_n$ is also the same as before.

The total number of quantum states for all permissible configurations is clearly $d(A)=\sum_{
s_{n,m}}d_{s_{n,m}}$. To estimate $d(A)$ we again expand $\ln d$ around the dominant
configuration (\ref{news}), denoted by $\bar s_{n,m}$. As before, it gives $\ln d=\ln d_{\bar
s_{n,m}}-{1\over 2}\sum\delta s_{n,m}K_{n,m;n'm'}\delta s_{n'm'}+o(\delta s_{n,m}^2)$ where
$K$ is the symmetric matrix $K_{n,m;n'm'}=\delta_{nn'}\delta_{mm'}/\bar
s_{n,m}-1/\sum_{k,l}\bar s_{k,l}$. All variations $\bar s_{n,m}+\delta s_{n,m}$ must satisfy the two conditions
(\ref{newc}) which give the two conditions $\sum\delta s_{n,m}\sqrt{n(n+2)}=0$ and
$\sum\delta s_{n,m}m=0$. Taking into account these equations 
the total number of states can be expressed as
\beq d&=&d_{\bar s_{n,m}}\sum_{-\infty}^\infty e^{-{1\over 2}\sum \delta
s_{n,m}K_{n,m;n'm'}\delta s_{n'm'}}\;\delta(\sum\delta s_{n,m}\sqrt{n(n+2)})\;
\delta(\sum\delta s_{n,m}m)\nn
&=&C'd_{\bar s_{n,m}}\Big[\prod_{n,m}\sqrt A\,\Big]/A,
\eeq
where $C'$ is again independent of $A$. Inserting (\ref{news}) into (\ref{newco}) and
dropping $o(1)$ factors we get 
\beq d_{\bar s_{n,m}}=\exp(\lambda A)(\sum\bar
s_{n,m})^{1/2}/\prod_{n,m}(2\pi\bar s_{n,m})^{1/2}.\eeq
Plugging these expressions into $d$ we finally get
\beq d={\alpha\over\sqrt A}\;e^{\lambda A}\;,\quad{\rm where}\;\alpha\sim o(1)\;,\eeq
leading once again to the formula (\ref{log}) for the entropy.
The origin of an extra $\sqrt A$ can be easily traced in this approach, which is the
additional condition $\sum ms_{n,m}=0$. Thus the coefficient of the log-correction is
absolutely robust and does not depend on the details of the configurations at all. It is
directly linked with the boundary conditions the horizon must satisfy.

%%%%%%%%%%%%%%%%%%%%%%
%\noindent{\bf Acknowledgments} :
%References

%%%%%%%%%%%%%%
\end{document}